\documentclass{article}
\usepackage{graphicx}
\title{\bf Kinks, energy conditions and closed timelike curves}
\author{ Pedro F. Gonz\'{a}lez-D\'{\i}az.\\
Instituto de Matem\'{a}ticas y F\'{\i}sica Fundamental\\ Consejo Superior
de Investigaciones Cient\'{\i}ficas\\ Serrano 121, 28006 Madrid,
SPAIN\\ }
\date{April 7, 2000}
\begin{document}
\maketitle
\large
\setlength{\baselineskip}{0.5cm}

\begin{abstract}
A link between the possibility of extending a geodesically
incomplete kinked spacetime to a spacetime which is geodesically
complete and the energy conditions is discussed for the case of
a cylindrically-symmetric spacetime kink. It is concluded that
neither the strong nor the weak energy condition can be
satisfied in the four-dimensional example, though the latter
condition may survive on the transversal sections of such a
spacetime. It is also shown that the matter which propagates
quantum-mechanically in a kinked spacetime can always be trapped
by closed timelike curves, but signaling connections between
that matter and any possible observer can only be made of
totally incoherent radiation, so preventing observation of
causality violation.

\end{abstract}

\renewcommand{\theequation}{\arabic{section}.\arabic{equation}}

\pagebreak

\section{\bf Introduction}

Chamblin has recently stressed$^{1}$ that the possibility of
extending a non-spacelike geodesically incomplete kinked
spacetime to a spacetime that is geodesically complete hinges on
the energy conditions being satisfied. On the other hand,
Gibbons and Hawking proposed$^{2}$ that kinked space-times
should be related to the existence of closed timelike curves
(CTCs). This proposal was subsequently critisized by Chamblin
and Penrose$^{3}$, who showed that this relation could not be
established, at least for classical space-times and matter.

The above alluded connections have been already discussed in the
case of de Sitter space kink$^{4}$ and other
spherically-symmetric gravitational topological defects$^{5,6}$.
It turned out that in these examples the energy conditions were
not satisfied$^{7}$ and, however, they all can be maximally
extended to a geodesically complete spacetime. Moreover, none of
these space-times may contain CTCs when their matter contents
are considered classically$^{4,8}$, in accordance with the
Chamblin-Penrose claim$^{3}$.

As a first step to investigate the influence of space-time
symmetry on these issues, they will be discussed in the present
paper for the case that the kink occurs in a
cylindrically-symmetric space-time.

\section{\bf Cylindrically-symmetric kink}
\setcounter{equation}{0}

The typical example of a cylindrically-symmetric kink is the
kink in the interior of an extreme cosmic string$^{9,10}$. This
is characterized by a gravitational coupling $G\mu=\frac{1}{2}$,
where $\mu$ is the string linear density. In what follows, let
us briefly first review the topological properties of the kinked
extreme string, and then comment on some aspects of its
geometry.

The general concept of a gravitational kink can be introduced by
starting with the Lorentz metric $g_{ab}$ of a four-dimensional
spacetime as given by a map, $P$, from any connected
three-manifold, $\partial${\bf M}, of the spacetime
four-manifold, {\bf M}, into the set of timelike directions in
{\bf M} $^{11}$. Metric homotopy can then be classified by the
degree of this map, and the kink number (or topological charge)
of the Lorentz metric, with respect to a hypersurface $\Sigma$,
can be defined by$^{11}$
\[{\rm Kink}(\Sigma;g_{ab})={\rm deg}(P),\]
so that the gravitational kink can be viewed as a measure of how
many times the light cones rotate around as one moves along
hypersurface $\Sigma$.

In the case of the spacetime of an extreme cosmic string, whose
interior geometry can be visualized as that of a sphere when the
corresponding two-metric is embedded in an Euclidean
three-sphere$^{12}$, the pair ($\Sigma;g$) will describe a
gravitational kink with topological charge $\kappa=+1$ if ${\rm
Kink}(\Sigma;g)=1$. From the above discussion, one may also
visualize the internal geometry of the extreme string by
enforcing the constant-time sections, $\tau=\tau_{0}$, of the
interior metric of a string$^{9,12}$ with uniform density
$\epsilon$, out to some cylindrical radius,
\begin{equation}
ds^2=\frac{dr^2}{1-\frac{r^2}{r_{*}^2}}+dz^2+r^2d\phi^2,
\end{equation}
where
\begin{equation}
r=r_{*}\sin\frac{\rho}{r_{*}},
\end{equation}
with $r_{*}=(8\pi G\epsilon)^{-\frac{1}{2}}$, and
$-\infty<z<\infty, 0\leq\phi\leq2\pi, 0\leq\rho\leq
r_{*}\arccos(1-4G\mu)$, to be isometrically embedded in the
kinked spacetime. The corresponding cylindrically-symmetric
standard, kinked metric is given by$^{9,13}$
\begin{equation}
ds^2=-\cos 2\alpha d\hat{t}^2\mp 2kd\hat{t}dr+dz^2+r^2d\phi^2,
\end{equation}
where the upper/lower sign of the second term corresponds to a
positive/negative topological charge, $k=\pm 1$, depending on
which of the two coordinate patches required for a complete
description of the kink is being considered$^{9}$, and $\alpha$
is the tilt angle of the light cones in the kink,
$0\leq\alpha\leq\pi$. The isometric embedding will hold if in
metric (2.3) we have furthermore
\begin{equation}
\cos 2\alpha=1-\frac{r^2}{r_{*}^2}
\end{equation}
and
\begin{equation}
\hat{t}= \tau_{0}-k\int\frac{dr}{\cos 2\alpha}.
\end{equation}

Actually, a gravitational kink depends only on D-1 of the D
spacetime coordinates, and is spherically symmetric on them.
However, the cylindric coordinate $z$ in metric (2.1) and (2.3)
is not going to play any role in the analysis to follow and,
therefore, one could reduce these metrics just to their
hemispherical $z$=const. sections. On the other hand, one can
also embed the $z$=const. sections of metric (2.3) in an
Euclidean space and, hence re-express that metric in an explicit
spherically-symmetric form:
\[ds^2=-\cos 2\alpha d\hat{t}^2-kd\hat{t}dr+r_{*}^2d\Omega_{2}^2,\]
where we have specialized to the case of a gravitational kink
with positive topological charge, $\kappa=+1$,
$d\Omega_2^2=d\theta^2 +\sin^2\theta d\phi^2$ is the metric on
the unit two-sphere, and we have used Eq. (2.5).

The metric (2.3) can be then regarded as the metric for the
embedding of metric (2.1), and the kinked time $\hat{t}$, as the
corresponding embedding function. Hence, one can obtain an
embedding ``rate''
\begin{equation}
\frac{d^2 r}{d\hat{t}^2}=\frac{2r}{r_{*}}\left(\frac{r^2}{r_{*}^2}-1\right),
\end{equation}
which tells us that the embedding surface would flare either
outward if $r<r_{*}$, or inward if $r>r_{*}$. The string metric
(2.1) should now be interpreted as a kinked boundary in the
space with kinked spacetime (2.3).

If the isometric embedding of metric (2.1) in metric (2.3)
holds, from (2.2) and (2.4) we have $\cos^2\theta=\cos 2\alpha$,
with $\theta=\frac{\rho}{r_{*}}$, and if the one-kink is
conserved, then $G\mu$ is enforced to be $\frac{1}{2}$ and $r$
should be analytically continued beyond $r_{*}$, up to
$\sqrt{2}r_{*}$$^{9}$. This extension creates a spherical shell
filled with broken phase at each $z$-const. section, preventing
the extreme string with $G\mu=\frac{1}{2}$ from disappearing,
and converts the conical singularity at $r=r_{*}$ into a de
Sitter-like cosmological singularity (horizon)$^{9}$. All of the
topological charge of the kink would then be confined within the
shell, that is within a finite compact region beyond the
cosmological horizon that extends up to $r=\sqrt{2}r_{*}$.
Inside the horizon all hypersurfaces $\Sigma$ are everywhere
spacelike. Thus, as a consequence from the back reaction of the
gravitational field of the one-kink, the lost picture of a
cosmic string with a core region of trapped energy would be
recovered for the extreme string with $G\mu=\frac{1}{2}$.

We have established a consistent and regular embedding of the
extreme string metric in each of the two patches of the kinked
spacetime whose surfaces would, according to expression (2.6),
flare outward at $\sqrt{2}r_{*}$, with a maximum ``rate''
\[\left.\frac{d^2 r}{d\hat{t}^2}\right|_{r=\sqrt{2}r_{*}}=\frac{2\sqrt{2}}{r_{*}}.\]

To stationary observers at the center of the sphere
corresponding to each surface $z$=const., $\tau$=const., the
compact shell containing all the topological charge of the
kink$^{14}$ locally coincides with a finite region of the
exterior of either a de Sitter space when the light cones rotate
away from the observers (positive topological charge), or the
time-reverse to de Sitter space if the observers see light cones
ratating in the opposite direction (negative topological
charge). In the latter case, only the region outside the
cosmological horizon would be accessible to stationary
observers.

The kink metric (2.3) is geodesically incomplete as it shows an
apparent horizon at $r=r_{*}$ on the two coordinate patches.
However, it can be converted into a geodesically complete metric
by introducing Kruskal coordinates. It was obtained in Ref. 9
that the maximally-extended metric describing the spacetime of
an extreme string kink can be written as
\begin{equation}
ds^2=-\frac{4kr_{*}^2}{(k-UV)^2}dUdV+dz^2+r^2d\phi^2,
\end{equation}
where again $k(=\pm 1)$ labels the two coordinate patches
required to describe a complete one-kink, and $U$ and $V$ are
the Kruskal coordinates$^{9}$
\begin{equation}
U=\mp
e^{-\frac{k\hat{t}}{r_{*}}}\left(\frac{r_{*}-r}{r+r_{*}}\right),\;\;
V=\pm e^{\frac{k\hat{t}}{r_{*}}},
\end{equation}
in terms of which the radial coordinate can be defined as
\begin{equation}
r=r_{*}\left(\frac{k+UV}{k-UV}\right),
\end{equation}
with the time $\hat{t}$ given by
\[\hat{t}=t
-kr_{*}\sqrt{2\left(1-\frac{r^2}{2r_{*}^2}\right)}\]
\begin{equation}
+\frac{1}{2}kr_{*}\ln\left[\frac{\left(1+
\sqrt{4\left(1-\frac{r^2}{2r_{*}^2}\right)}\right)(r-r_{*})}
{\left(1-
\sqrt{4\left(1-\frac{r^2}{2r_{*}^2}\right)}\right)(r+r_{*})}\right],
\end{equation}
where $t$ is the metrical kinked time which is related to the
time entering the metric of the kinkless cosmic string$^{9}$.

The interesting feature of metric (2.7) is that its $z$-const.
sections are exactly the metric which describes a hemispherical
section of the de Sitter spacetime kink$^{4}$ for a positive
cosmological constant $\Lambda=\frac{3}{r_{*}^2}$.

\section{\bf Energy conditions in cylindrically-symmetric kinks}
\setcounter{equation}{0}

The simplest, general metric describing the spacetime of a
cylindrically-symmetric kink can be written as
\begin{equation}
ds^2=-\cos 2\alpha(dt^2-dr^2)-2\sin 2\alpha dtdr+dz^2+r^2
d\phi^2.
\end{equation}
In order to investigate the possibility that geodesically
incomplete kinks can be extended to geodesically complete ones
when the energy conditions are satisfied in a
cylindrically-symmetric space-time, let us review these
conditions and also physical conditions of the kind considered
by Finkelstein and McCollum$^{13}$, by using the Hawking-Ellis
procedure$^{15}$. Thus, we write the eigenvalue equation
\begin{equation}
\left(G_{\mu\nu}-\lambda g_{\mu\nu}\right)\xi^{\nu}=0.
\end{equation}
For this equation to be implemented in the case of a
cylindrically-symmetric space-time kink, we obtain first the
nonzero Chrystoffel symbols for metrics (3.1). These are:
\[\Gamma_{tt}^t=\sin^2 2\alpha\partial_r \alpha,\;\;
\Gamma_{tr}^t =\Gamma_{tt}^r =-\sin 2\alpha\cos 2\alpha\partial_r \alpha,\]

\[\Gamma_{rr}^{t}=(1+\cos^{2}2\alpha)\partial_{r}\alpha,\;\;
\Gamma_{tr}^{r}=-\sin^{2}2\alpha\partial_{r}\alpha,\;\;\]
\begin{equation}
\Gamma_{\phi\phi}^{t}=r\sin 2\alpha,
\end{equation}
\[\Gamma_{rr}^{r}=\sin2\alpha\cos2\alpha\partial_{r}\alpha,\;\;
\Gamma_{r\phi}^{\phi}=r^{-1},\;\; \Gamma_{\phi\phi}^{r}=-r\cos2\alpha.\]
Hence we can obtain the nonzero components of the Ricci tensor
which are:
\[R_{tt}=-R_{rr}=-\frac{1}{2r}\cos2\alpha\partial_{r}(r^{2}\triangle),\]
\begin{equation}
R_{tr}=-\frac{1}{2r}\sin2\alpha\partial_{r}(r^{2}\triangle),
\end{equation}
\[R_{\phi\phi}=r^{2}\triangle ,\]
where
\begin{equation}
\triangle=\frac{2}{r^{2}}\partial_{r}(r\sin^{2}\alpha)
=\frac{2}{r^{2}}\partial_{r}\mu,
\end{equation}
with $\mu$ ($=r\sin^{2}\alpha$) the function introduced by
Finkelstein and McCollum$^{13}$. For the curvature scalar we
then obtain
\begin{equation}
R=\triangle+\frac{1}{r}\partial_{r}(r^{2}\triangle),
\end{equation}
and for the nonzero mixed components of the Einstein tensor
\[G_{t}^{t}=G_{r}^{r}=-\frac{1}{2}\triangle=-\frac{1}{r^{2}}\partial_{r}\mu\]
\begin{equation}
G_{z}^{z}=-\frac{1}{r^{2}}\partial_{r}\mu-\frac{1}{r}\partial_{r}^{2}\mu
\end{equation}
\[G_{\phi}^{\phi}=\frac{1}{r^{2}}\partial_{r}\mu-\frac{1}{r}\partial_{r}^{2}\mu.\]
Introducing these Einstein-tensor components in the eigenvalue
equation (3.2), we obtain the eigenvalues
\[\lambda_0=\lambda_1=\frac{1}{r^{2}}\partial_{r}\mu\]
\begin{equation}
\lambda_2=-\lambda_1=\frac{1}{r^{2}}\partial_{r}\mu-\frac{1}{r}\partial_{r}^{2}\mu
\end{equation}
\[\lambda_3=\lambda_1=\frac{1}{r^{2}}\partial_{r}\mu-\frac{1}{r}\partial_{r}^{2}\mu.\]
The corresponding eigenvectors are
\[E_0=\left(\cos\alpha, \sin\alpha, 0, 0\right)\]
\[E_1=\left(\sin\alpha, -\cos\alpha, 0, 0\right)\]
\begin{equation}
E_2=\left(0, 0, 1, 0\right)
\end{equation}
\[E_3=\left(0, 0, 0, \frac{1}{r}\right).\]

Since $E_0$ is timelike and the $E_{\rho}$'s ($\rho=1,2,3$) are
all spacelike, we have a canonical form of Type I, according to
the classification of Hawking and Ellis$^{15}$. The spacelike
eigenvectors $\{E_{\rho}\}$ form an orthonormal basis and the
tetrad components of the metric tensor are
\begin{equation}
\bar{g}_{\rho\sigma}=g(E_{\rho},E_{\sigma})={\rm diag}(-1,1,1,1).
\end{equation}
Finally, we obtain for the tetrad components of the
energy-momentum tensor
\begin{equation}
\|\bar{T}^{\rho\sigma}\|={\rm diag}(\epsilon,p_1,p_2,p_3),
\end{equation}
with
\begin{equation}
\epsilon=\frac{1}{r^2}\partial_{r}\mu,\;\;p_1=-\frac{1}{r^2}\partial_{r}\mu,
\end{equation}
\begin{equation}
p_2=-\frac{1}{r^2}\partial_{r}\mu-\frac{1}{r}\partial_{r}^{2}\mu,
\end{equation}
\begin{equation}
p_3=\frac{1}{r^2}\partial_{r}\mu-\frac{1}{r}\partial_{r}^{2}\mu.
\end{equation}

We can now discuss the energy conditions in general
cylindrically-symmetric kinked space-times. The weak energy
condition states$^{15,16}$ that the energy density as measured
by an observer must be non-negative, and this requires
\begin{equation}
\epsilon\geq 0,\;\; \epsilon+p_{\rho}\geq 0, \;\; \rho=1,2,3 .
\end{equation}
Using (3.12)-(3.14), inequalities (3.15) lead to:
\begin{equation}
\partial_{r}\mu\geq 0
\end{equation}
\begin{equation}
\partial_{r}\left(\frac{1}{r^{2}}\partial_{r}\mu\right)\leq 0
\end{equation}
\begin{equation}
\partial_{r}^{2}\mu\leq 0.
\end{equation}

On the other hand, for a canonical form of type I there also
holds the strong energy condition$^{15,16}$, provided that
\begin{equation}
\epsilon+p_{\rho}\geq 0,\;\; \epsilon+\sum_{\rho}p_{\rho}\geq 0,
\;\; \rho=1,2,3 .
\end{equation}
For (3.12)-(3.14) these inequalities lead to (3.18) again. We
note that although the strong energy condition does not imply
the weak energy condition, the vice versa is true however.

Physical conditions which function $\mu(r)$ must satisfy$^{7}$
are: (a) $\partial_{r}\mu\geq 0$ for all $r$, (b)
$\partial_{r}\left(\frac{1}{r^{2}}\partial_{r}\mu\right)\leq 0$
for all $r$, (c) $\partial_{r}^{2}\mu\leq 0$ for all $r$, (d)
$\mu=O(r^{3})$ as $r\rightarrow 0$ in order for
$|G_{\rho}^{\sigma}| < \infty$ at $r=0$, and (e)
$0\leq\frac{\mu}{r}\leq 1$ in order for $0\leq\sin^{2}\alpha\leq
1$. On the other hand, to ensure the existence of an one-kink,
we need to impose the boundary conditions $\alpha(0)=0$ and
$\alpha(0)=\pi$, since in the present case,
$\alpha=\arcsin\left(\frac{r}{\sqrt{2}r_{*}}\right)$. Hence,
recalling that $\frac{\mu}{r}=\sin^{2}\alpha$, we have the
additional condition (f) $\lim_{r\rightarrow 0}\frac{\mu}{r}=0$.
From this condition it follows $\mu=O(r^{1+c})$, with $c>0$ for
small $r$, and hence $\partial_{r}\mu=O(r^{c})>0$ and
$\partial_{r}^{2}\mu= O(r^{c-1})>0$, for small $r>0$. The latter
inequality violates condition (c) and, therefore, not only the
strong energy condition but also the weak energy condition is
violated in this space-time. Thus, like for kinked de Sitter
space, space-time kinks possessing cylindrical symmetry should
violate the above two energy conditions, in spite of they being
maximally extendible to geodesically complete space-time.
Moreover, it can be readily seen that these violations also
imply violation of the dominant energy condition$^{16}$,
$\epsilon\geq 0$, $p_{\rho}\in[-\epsilon,+\epsilon]$, and the
null energy condition$^{16}$ $\epsilon+p_{\rho}\geq 0$.

Let us finally consider what happens to the energy conditions
when we restrict ourselves to a $z=$const. section of metric
(3.1). It is easy to see then that the weak energy condition
implies only inequalities (3.16) and (3.17) to hold, while the
strong energy condition leads to (3.17) again. It follows that
the former condition would only be satisfied if $0\leq c\leq 2$,
while the latter one holds when $0\leq c\leq 1$. Since in the
actual case $c=2$, we see that for $z$=const. sections of metric
(3.1), the weak energy condition is satisfied but the strong
energy condition remains being violated. This conclusion can be
extended to the geodesically complete space-times. Thus, the
Chamblin's argument$^{1}$ for cylindrically-symmetric kinks
should be relaxed so that a non-spacelike geodesically
incomplete kinked spacetime would be extendible to a
geodesically complete one if the weak (but not the strong)
energy condition is preserved on their $z=$const. sections.

\section{\bf Kinks and closed timelike curves}
\setcounter{equation}{0}

We shall analyse next the possible connection between kinks and
CTCs, suggested by Gibbons and Hawking$^{2}$, for the case of an
extreme cosmic-string kink. Both the discussion and conclusion
to be obtained can straightforwardly be generalized to any other
spacetime kinks.

The existence of CTCs in a given spacetime can be investigated
by considering the paths followed by null geodesics on the
Kruskal diagrams. For the extreme string kink with Kruskal
diagrams given in Fig. 1, it might seem at first glance that
since the two coordinate patches are identified on the surfaces
$r=A=\sqrt{2}r_{*}$, both on the original regions I and II and
the new regions III and IV (created by Kruskal extension),
because of continuity of the tilt angle at
$\alpha=\frac{\pi}{2}$, one could choose null geodesics that
started at $r=0$ in region I$_{+}$ and would somewhat loop back
through the new regions to finally arrive at their starting
point, after traversing both coordinate patches. However, one
can easily convince oneself that such itineraries are
classically disallowed, since they would require identification
of the two patches also on minimal surfaces at $r=0$ belonging
to a physical and a nonphysical region, respectively (see Fig.
1).

Nevertheless, the above conclusion is no longer valid when we
consider propagation of quantum fields in the same kinked
spacetime. The semiclassical regime for kinked spacetimes with
event horizons can simply be achieved by considering the
mathematical implications imposed by the fact that time
$\bar{t}$ enters the Kruskal coordinates $U,V$ in the form of
the dimensionless exponent $k\hat{t}/r_{*}$. The argument of the
logarithm in (2.10) becomes then square rooted and, therefore,
the expression for the time $\hat{t}$ entering the definition of
coordinates $U,V$ should be generalized to$^{9}$:
\begin{equation}
\hat{t}\rightarrow\hat{t}_{g}=\hat{t}+\frac{i}{2}k\kappa(1-\kappa)\pi r_{*},
\end{equation}
where $\hat{t}$ is given by (2.10) and $\kappa=\pm 1$. For
$\kappa=+1$, $\hat{t}_{g}=\hat{t}$, and for $\kappa=-1$, the
points ($\hat{t}-ik\pi r_{*},r,z,\phi$) in each patch are
actually the points in the same patch obtained by reflection in
the bifurcation point $U,V=0$, keeping the Kruskal metric real
and unchanged.

\begin{figure}[h]
\begin{center}
\includegraphics[width=.8\textwidth]{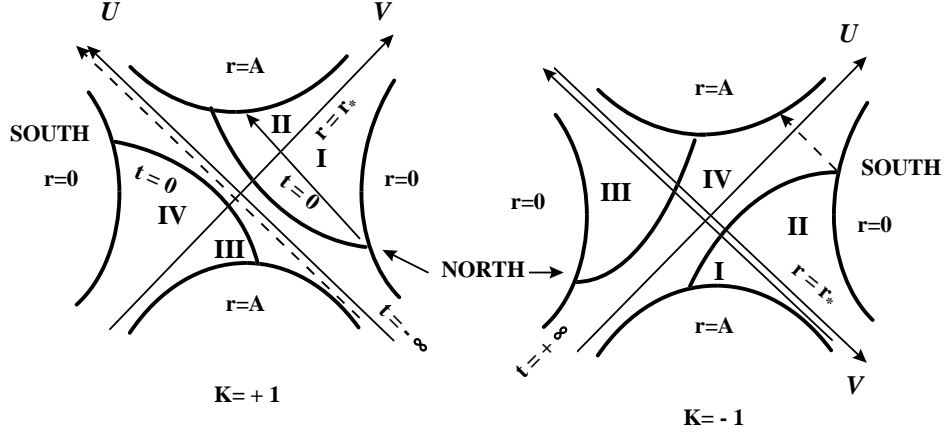}
\end{center}
\caption{Kruskal diagrams for the two coordinate patches of the one-kink
cosmic-string spacetime. The trajectories for some classical
null geodesics are shown as straight continuous or dashed
lines.Also shown are the geodesic trajectories at time $t=0$.
Pole identifications on the figure are arbitrary. }
\end{figure}

We note that one can still recover the standard kink metric
(2.3) from a general Kruskal metric if we redefine the Kruskal
coordinates as follows:
\begin{equation}
U=\pm 2b\kappa
e^{-\frac{k\hat{t}_{c}}{r_{*}}}\left(\frac{r_{*}-r}{r_{*}+r}\right)
\end{equation}
\begin{equation}
V=\mp k\kappa e^{\frac{k\hat{t}_{c}}{r_{*}}},
\end{equation}
where
\begin{equation}
\hat{t}_{c}=\hat{t}+i\pi r_{*}\kappa k.
\end{equation}

This choice leaves the expressions for $UV, F, r$ and the
Kruskal metric real and unchanged. For $\kappa=-1$, Eqns. (2.8)
become the sign-reversed of (4.2) and (4.3), respectively; i.e.:
the points ($\hat{t}-ik\pi r_{*},r,z,\phi$) on the Kruskal
diagrams of the two coordinate patches are the points in the new
regions III$_{k}$ on the same diagrams, obtained by reflecting
in the origins of the respective $U,V$ planes, preserving the
Kruskal metric real and unchanged. This leads to identification
of hyperbolae in the new regions with hyperbolae in the original
regions for the same values of $r$; i.e. to identification of
hyperbolas III$_{k}$ with hyperbolas II$_{k}$ and hyperbolas
IV$_{k}$ with hyperbolas I$_{k}$. We note that the existence of
such identifications in turn amounts to both the kind of
periodicity required by Hawking thermal radiation$^{17}$ in each
patch, and the existence of CTCs.

Since in the semiclassical description, we can identify maximum
surfaces of the physical regions with those of the nonphysical
regions in each coordinate patch separately, we recover
allowance for the null geodesics that start at surface $r=0$ in
region I$_{+}$ of patch $k=+1$ (Fig. 1) to continue propagating
on patch $k=-1$, after the maximum surface of region III$_{-}$,
first through region II$_{-}$ and then through region IV$_{-}$,
up to the surface at $r=A=\sqrt{2}r_{*}$ of the latter new
region. Because this surface can be identified with the similar
surface in region III$_{+}$ of patch $k=+1$, the considered null
geodesics can thereafter propagate into the region IV$_{+}$ and,
again by quantum identification of surfaces at $r=0$, come back
to their starting points on the surface at $r=0$ of region
I$_{+}$, in patch $k=+1$. Hence, null geodesics starting from
original regions at $r=0$ can still loop back to arrive at their
starting points, so completing a CTC, provided such a CTC is
involved at a thermal radiation process preventing any
information to flow from or to the CTC. Our general conclusion
therefore is that CTCs are linked to spacetime kinks if and only
if the matter traveling through these spacetimes is considered
quantum-mechanically.

\section{\bf Conclusion}

The problem of the relation between geodesic incompleteness and
the holding of the energy conditions has been considered for the
case of a kink in a cylindrically-symmetric spacetime. We
discussed the physical conditions that such a spacetime must
satisfy in relation with the weak and strong energy conditions.
It was shown that, whereas none of these conditions holds for
the four-dimensional case, the weak energy condition can still
survive on the $z$ = const. (transversal) sections of the
spacetime. Also considered has been the problem of the
connection between spacetime kinks and the existence of closed
timelike curves. We obtained that kinked spacetimes do not
contain CTCs if the matter propagating on them is dealt with
classically, but CTCs become a necessary ingredient in such
spacetimes whenever the propagating matter shows quantum
behaviour. As it was pointed out first in Ref. [18] and later in
Ref. [19] there must be a close connection between CTCs and the
thermal processes induced by the presence of an event horizon.
Any possible observer of the quantum matter trapped in a CTC
could only detect it by means of a totally incoherent radiation
carrying no desciphrable information to or from the observer.
Thus, the connection of CTCs with the thermal process prevents
the existence of any observable causality violating process
induced by the CTCs and, therefore, allows one to conjecture a
censorship for causality violation, even when CTCs and time
machines can exist and be operative. One might illustrate the
resulting situation by re-paraphrasing Stephen Hawking [30]:
there could perfectly be hords of tourists visiting us from the
future, but neither they nor we could know anything about their
trip. For them, it would be a touring which costs a lot and
rewards nothing; for us, the trip would simply be unnoticed.

\vspace{.8cm}

\noindent{\bf Acknowledgements}

The author acknowledges DGICYT by support under Research Project
No. PB97-1218

\noindent\section*{References}
\begin{description}
\item [1] A. Chamblin, J. Geom. Phys. 13, 357 (1994).
\item [2] G.W. Gibbons and S.W. Hawking, Phys. Rev. Lett.
69, 1719 (1992).
\item [3] A. Chamblin and R. Penrose, Twistor Newsletter 34, 13 (1992).
\item [4] K.A. Dunn, T.A. Harriott and J.G. Williams, J. Math. Phys.
35, 4145 (1994).
\item [5] P.F. Gonz\'{a}lez-D\'{\i}az, Phys. Rev. D51, 7144 (1995).
\item [6] P.F. Gonz\'{a}lez-D\'{\i}az, in: {\it Theories of Fundamental
Interactions; Maynooth Bicentenary Volume}, ed. T. Tchrakian
(World Scientific, Singapore, 1995); Grav. Cosm. 2, 621 (1996).
\item [7] K.A. Dunn, T.A. Harriott and J.G. Williams, J. Math. Phys.
37, 5637 (1996).
\item [8] P.F. Gonz\'{a}lez-D\'{\i}az, Int. J. Mod. Phys. D7, 793 (1998).
\item [9] P.F. Gonz\'{a}lez-D\'{\i}az, Phys. Rev. D52, 5698 (1995).
\item [10] A. Vilenkin and E.P.S. Shellard, {\it Cosmic Strings
and other Topological Defects} (Cambridge U. P., Cambridge, UK,
1994).
\item [11] G.W. Gibbons and S.W. Hawking, Commun. Math. Phys. 142, 1 (1990).
\item [12] J.R. Gott, Astrophys. J. 288, 422 (1985); W.A. Hiscock,
Phys. Rev. D31, 3288 (1985).
\item [13] D. Finkelstein and C.W. Misner, Ann. Phys. (N.Y.) 6, 230 (1959).
\item [14] A. Chamblin, {\it Kinks and singularities}, Preprint,
DAMTP-R95/44 (1995).
\item [15] S.W. Hawking and G.F.R. Ellis, {\it The Large Scale
Structure of Space-Time} (Cambridge U. P., Cambridge, UK, 1973).
\item [16] M. Visser, {\it Lorentzian Wormholes} (American I. P. P.,
Woodbury, USA, 1996).
\item [17] J.B. Hartle and S.W. Hawking, Phys. Rev. D31, 2188 (1976).
\item [18] P.F. Gonz\'{a}lez-D\'{\i}az, {\it Time Walk through the
quantum cosmic string}, gr-qc/9412034.
\item [19] S.W. Hawking, Phys. Rev. D52, 5681 (1995).
\item [20] S.W. Hawking, Phys. Rev. D46, 603 (1992).
\end{description}

\end{document}